\documentclass[sigconf]{acmart}

\setcopyright{none}

\acmConference[WebSci '20]{12th ACM Conference on Web Science}{July 6--10, 2020}{Southampton, United Kingdom}

\usepackage{graphicx,url}
\usepackage[utf8]{inputenc}
\usepackage{graphicx}
\usepackage{caption}

\usepackage{todonotes}
\usepackage{multirow}
\usepackage{textcomp, color, soul}
\usepackage{tabularx}
\usepackage{url}
\usepackage{subcaption}
\usepackage{float}
\usepackage{amsmath}
\usepackage{booktabs}
\usepackage{balance}
\usepackage{courier}

\begin{document}

\title[How Biased is the Population of Facebook Users?]{How Biased is the Population of Facebook Users? Comparing the Demographics of Facebook Users with Census Data to Generate Correction Factors}


\author{Filipe N. Ribeiro}
\affiliation{%
  \institution{Universidade Federal de Ouro Preto}
  \city{Ouro Preto} 
  \state{Brazil} 
}
\email{filipe.ribeiro@ufop.edu.br}

\author{Fabrício Benevenuto}
\affiliation{%
  \institution{Universidade Federal de Minas Gerais}
  \city{Belo Horizonte} 
  \state{Brazil} 
}
\email{fabricio@dcc.ufmg.br}

\author{Emilio Zagheni}
\affiliation{%
  \institution{Max Planck Institute for Demographic Research}
  \city{Rostock} 
  \state{Germany} 
}
\email{zagheni@demogr.mpg.de}

\keywords{social media, advertising, census}
\begin{abstract}
Censuses and representative sampling surveys around the world are key sources of data to guide government investments and public policies. However, these sources are very expensive to obtain and are collected relatively infrequently. Over the last decade, there has been growing interest in the use of data from social media to complement more traditional data sources. However, social media users are not representative of the general population. Thus, analyses based on social media data require statistical adjustments, like post-stratification, in order to remove the bias and make solid statistical claims. These adjustments are possible only when we have information about the frequency of demographic groups using social media. These data, when compared with official statistics, enable researchers to produce appropriate statistical correction factors. In this paper, we leverage the Facebook advertising platform to compile the equivalent of an aggregate-level census of Facebook users. Our compilation includes the population distribution for seven demographic attributes such as gender, political leaning, and educational attainment at different geographic levels for the U.S. (country, state, and city). By comparing the Facebook counts with official reports provided by the U.S. Census and Gallup, we found very high correlations, especially for political leaning and race. We also identified instances where official statistics may be underestimating population counts as in the case of immigration. We use the information collected to calculate bias correction factors for all computed attributes in order to evaluate the extent to which different demographic groups are more or less represented on Facebook, and to derive the actual distributions for specific audiences of interest. We provide the first comprehensive analysis for assessing biases in Facebook users across several dimensions. This information can be used to generate bias-adjusted population estimates and demographic counts in a timely way and at fine geographic granularity in between data releases of official statistics.



\end{abstract}

\maketitle

\section{Introduction}

Censuses have been used for many centuries to assess demographic quantities. They are necessary and of utmost importance for the orderly functioning of modern societies. Censuses are crucial for defining priority investments for education, infrastructure and other public policies. In countries like the US, data collection through censuses is mandated by the Constitution. Censuses are necessary; however, the cost and time needed to run a census of the population are quite high. A recent report published by the U.S. Census Bureau estimates that the expected cost for the 2020 decennial Census is 15 billion dollars \footnote{\url{https://www2.census.gov/programs-surveys/decennial/2020/program-management/planning-docs/2020-cost-estimate1.pdf}}.


Complementary forms of data collection for censuses have been tested by different countries. In Norway, for instance, authorities conducted the Census with a register-based approach, which uses information from an existing administrative source and gather information about households, dwellings and individuals to complement data about the population's demographic characteristics. This technique depends on a unique identification number across different administrative sources and may not be used in all countries also because of legal restrictions related to using these data for the purposes of  statistical analysis. An alternative, tested by Spain, uses both the register-based approach and the standard Census. France has tested an approach that relies upon collecting data in a cumulative survey that covers the country for years instead of a short period. In addition to this, researchers have proposed alternative/complementary approaches to infer demographic aspects from different sources.

In the context of social computing, inference of demographic features from the online world 
has received significant attention  since the early days of the World Wide Web (WWW). Back in $1997$, researchers developed the Lifestyle Finder \cite{Krulwich1997}, a fortune teller Web application that asked questions about demographic attributes, interests, and leisure activities to infer other demographic characteristics and recommend Web pages that the user would likely enjoy. Since its beginning, the WWW has experimented a huge growth in terms of number of users and variety of available services. In the same vein, collection of a large quantity of data about users has increased exponentially together with new possibilities to extract demographic information from online data.

The services and useful insights that can be leveraged using demographic data are not limited to recommending Web pages that fit user's profile.  Efforts in this area include studies that attempt to infer the political leaning of Online Social Network (OSN) users ~\cite{Truthy_icwsm2011politics,Golbeck:2011:CPP:1978942.1979106,Makazhanov:2013:PPP:2492517.2492527,10.1371/journal.pone.0137422}, and to detect gender to help forensic investigations~\cite{Vel2002}. 
In particular, some recent studies have explored OSN advertising platforms to infer demographics from aggregate information about users. These kinds of platforms rely on a rich source of data from users, such as workplace, visited venues, published posts, and `likes',  to infer users' demographic characteristics at a fine-grained level. 



In this paper, we gather estimates of demographic characteristics of Facebook users via the Facebook advertising platform, namely Facebook Ads. In particular, we analyze seven demographic categories collected through the advertising platform: gender, race, age, income, education, political leaning, and country of previous residence - and compare them to official statistics.


Our results show that part of the demographic data extracted from Facebook Ads is quite similar to official data, notably regarding race, political leaning, and graduate education level. For the categories where online data deviate from official statistics, we assess how much the online demographic groups are more or less represented on Facebook and we calculate correction factors.


By conducting this study, we intend to shed light on the rich demographic data amassed by OSN advertising services that might be much more useful for the academic community if systematically validated. 

As an additional output of this work, we release our data set with estimated correction factors for each demographic attribute. This enables one to generate estimates that approximate the original Census values, using a statistical adjustment procedure known as post-stratification. 


\section{Related Work}

In recent years we have witnessed large efforts in demographic research  to assess population characteristics from online environments~\cite{Cesare2018}. Researchers have used many sources of data to infer demographics such as email data~\cite{Zagheni:2012:YYE:2380718.2380764}, Google Plus~\cite{Messias2016@asonam}, and Twitter~\cite{mislove2011understanding,zagheni2014inferring}. In particular, several efforts have explored the OSN advertising platforms as a source of information. 


Advertisements underpin much of the Internet economy, and play a key role in the OSN business model. Consolidated multinational companies or even local small businesses around the world can take advantage of the ads infrastructure provided by OSN. With a global customer base, the revenues of companies like Facebook and Twitter have increased substantially and their market capital reached very high values. 
Not surprisingly, the Online Social Networks have revolutionized how advertisements are created and how to attract users' attention and engagement. Viral marketing techniques, close contact with customers, low costs,  and the possibility of targeting very specific niches of the population attracted advertisers from many different areas and sizes. The richness of data provided by OSN advertising platforms has been explored by the academic community to infer demographics across several research areas. 

Facebook Ads was used, for instance, to analyze the movement of migrants in the U.S.~\cite{zagheni2017leveraging} by counting the number of expats from 52 countries in the United States according to Facebook and comparing those values with data from the American Community Survey (ACS), a survey representative of the U.S. population, provided by the Census Bureau. The correlation found was very high even considering different age intervals and gender. A migration analysis extension work~\cite{Zagheni2018}, predicted migration of Mexicans to the U.S. by combining historical data from ACS and Facebook Ads data using a Bayesian hierarchical model. Other migration studies based on Facebook Ads shed light on the Venezuela's migration flow after the recent economic crisis ~\cite{10.1371/journal.pone.0229175} and the impact of Hurricane Maria on short-term mobility after the natural disaster~\cite{alexander2019impact}.

Facebook Ads was also used in health-related research, for monitoring countries with a high number of online users interested in lifestyle disease-related themes (diabetes, obesity, etc) and assessing how correlated that quantity is with the real prevalence of the respective diseases in the selected regions~\cite{araujo2017using,MejovaJMIR2018}. A related study examined the awareness of different demographic groups for topics related to schizophrenia on Facebook~\cite{info:doi/10.2196/jmir.6815} and found, for instance, that only 1\% of Facebook users in the United States have interests on schizophrenia-related themes and that women, those with lower education levels and Hispanics are more aware of this disease.

Gender gap studies were also conducted with data extracted from OSN advertising platforms. LinkedIn data was analysed to check the professional gender gaps across U.S. cities~\citep{professional_gender_gaps} whereas Facebook data was used to investigate the relationship in the gender gap verified in Facebook access and various indices of gender equality~\cite{Garcia6958,fatehkia2018using}. These studies confirmed the disparity of opportunities between genders and documented differences across countries.





Related approaches that rely on advertisement platforms were also employed to infer the political leaning of thousands of news media outlets in the U.S.~\cite{ribeiro2018@icwsm}, to detect the audience targeted by the socially divisive ads run on Facebook in the 2016 U.S. elections~\cite{ribeiro2019@fat}, to investigate the presence of LGBT communities across the U.S.~\cite{Gilroy2018}, and to measure cultural assimilation and distance across countries~\cite{dubois2018studying,stewart2019rock,carol_www20}.

The study that most approximates ours characterizes the population of Facebook users across 230 countries~\cite{8567877}. Although the main focus of this study is evaluating the growth of Facebook in terms of number of users and engagement, the authors collected the distribution of age and gender for each country they analyzed. 
Our effort focuses on assessing the similarities and differences between the demographic characteristics of OSN users and those of the underlying population. This is valuable because a better understanding of the attributes of online users would help us improve our understanding of population dynamics based on information for online users.

\section{Methodology}

The OSN advertising platforms provide basically three ways to define the audience that an ad should target. 1 - Personally Identifiable Information (PII) targeting, in which advertisers provide a list containing information that can link the customer with his/her online account such as email or phone number
; 2 - Look-alike audience targeting which is characterized by finding a similar audience to an initial set of customers, namely the source audience; 3 - Attribute-based targeting that allows the advertiser to define the targeted audience based on a range of attributes 
that include basic demographics (gender, age, and location); interests (entities in which user show an interest 
and can range from music preferences to religious orientations, or interest in artists and politicians); and behavioral characteristics such as `Business travelers' or `New vehicle buyers'
. Facebook, in particular, uses data provided by users when filling out their profile info (age, gender, educational level, and location) and infers other information by tracking user activities when using the platform or accessing external pages that use Facebook tools \footnote{\url{https://www.facebook.com/ads/about/?entry\_product=ad\_preferences}}.

We leverage the attribute-based targeting of Facebook Ads to infer demographics of U.S. users by exploring the combination of different attributes (targeting formula) and obtaining its audience size. 



As an example, take a simple targeting formula that selects all Facebook users that live in the U.S.. This formula includes people from both genders aged above 13 (minimum age allowed on Facebook) who lives in the United States, with an audience size of \textbf{$230$} millions of users\footnote{Data collected from Facebook marketing platform amassed data from Facebook and Instagram. Collection date: July, 2018.}. We can derive a new targeting formula in which we include a new attribute that limits the audience to only those people with a conservative political alignment (the corresponding attribute is `US politics (conservative)'). For this combination, the maximum number of users that can be reached is $39$ million. Likewise, we can replace the conservative-leaning attribute by liberal-leaning or very liberal-leaning attribute with an audience size of $47$ millions and $35$ millions of users, respectively. Finally, we can use the same strategy to assess the audience size for very conservatives and moderates, that represents $26$ millions and $45$ million, respectively.


\begin{figure*}[t!]
\centering
    \begin{subfigure}[b]{0.6\columnwidth}
    \centering
  \includegraphics[width=\columnwidth]{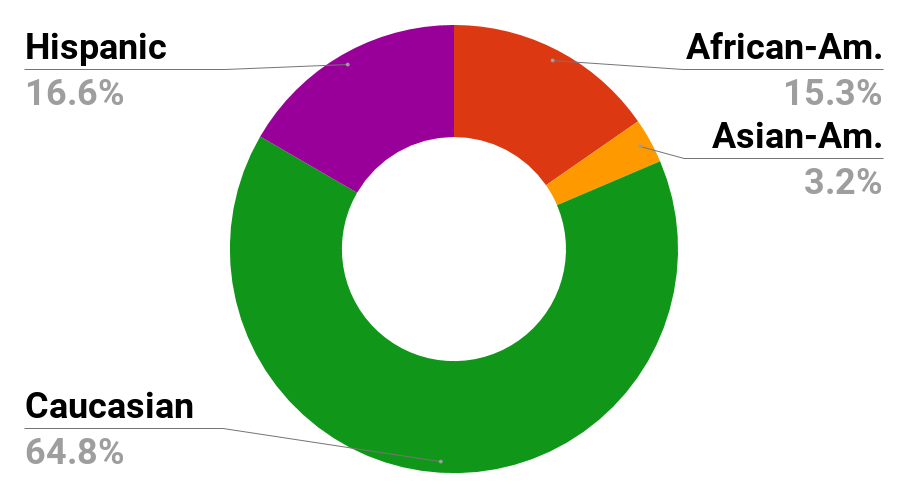}
    \caption{Race}
    \end{subfigure}\hfill
\begin{subfigure}[b]{0.6\columnwidth}
    \centering
  \includegraphics[width=\columnwidth]{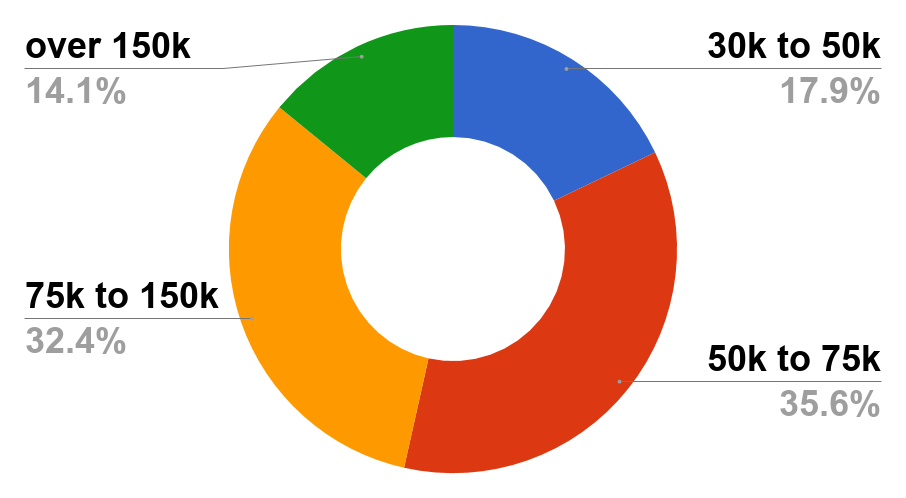}
    \caption{Income Level}
    \end{subfigure}\hfill
     \begin{subfigure}[b]{0.6\columnwidth}
     \centering
   \includegraphics[width=\columnwidth]{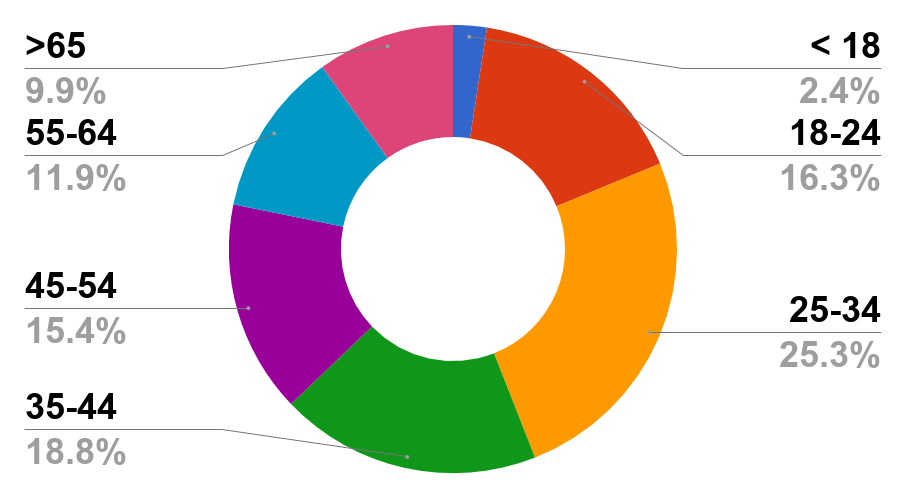}
     \caption{Age}
     \end{subfigure}
     \caption{\textbf{Demographic characteristics of U.S. Facebook users.}}
    \label{fig:framework_other_demograhpics}
\end{figure*}

Notice that from the initial targeting formula, we compute the amount of five subpopulations with different political alignments and based on this, we can derive the political leaning distribution for Facebook users who live in the U.S.. Similarly, we can also extend our initial targeting formula to infer the demographic attributes considered in the Census. Figure \ref{fig:framework_other_demograhpics} presents the distribution of three demographic attributes in the U.S. as extracted from Facebook Ads. We should mention that despite the absence of an attribute that identifies the most predominant race in the U.S., we compute this percentage by including in the targeting formula the negation for the Hispanic, African-American and Asian American.

In order to collect the audience sizes automatically, i.e. without the need to manipulate the UI, we used the Facebook Marketing API \footnote{\url{https://developers.facebook.com/docs/marketing-apis/}} that provides plenty of functions to help developers aiming at exploring Facebook advertising functions. 
In particular, it delivers a function called  `get reach estimate' that is key for our analysis. It allows developers to get the number of potential Facebook/Instagram users who match a specific targeting formula without the need to run an ad. 

In order to compare the Facebook Census with the actual population Census, we turn to the U.S. official authorities in this domain. The United States Census Bureau provides two annual reports in addition to the decennial Census. The ``American Community Survey'' (ACS) and the 
 ``Current Population Survey'' (CPS) are official surveys, curated by the official U.S. agency and have some significant differences in their methodologies\footnote{\url{https://www.census.gov/topics/income-poverty/poverty/guidance/data-sources/acs-vs-cps.html}}. ACS deals with a small number of indicators such as major income sources, however, the ACS data collection use a self-response mail questionnaire with an internet response option and with mandatory response, similar to the decennial census form. Conversely, the CPS provides much more detailed data including more comprehensive coverage of all potential income sources, but the data collection is conducted by interviewers via Computer Assisted Telephone Interviewing and  participation is not mandatory. In order to avoid issues with small sample sizes, we used the 2013-2017 ACS 5-year Estimates (ACS 2017)
 , released on December 8, 2018\footnote{\url{https://www.census.gov/programs-surveys/acs/news/data-releases/2017/release.html}}.
 
  
More specifically, we used the following ACS tables to obtain the official Census demographic data: S0101(age and gender), DP05(race), S2001(income), and S1501(education attainment), B05006 (immigrants). For the political leaning attribute we used a Gallup study based on party affiliation by state\footnote{\url{https://news.gallup.com/poll/226643/2017-party-affiliation-state.aspx}} as the baseline, since the Census do not include this attribute in their reports. 

All data obtained from the original Census were collected in three granularity levels: country, state, and city level. We collected the demographic distribution for the 50 most populated cities in the U.S. to provide a comparison in a more fine-grained level. In order to compare Facebook Ads data with the  Census data (for simplicity, we refer to ACS data as Census data, even though they are not the decennial Census) we calculated the Pearson correlation to check the linear correlation between each one of the demographic dimensions. 

A critical challenge in this analysis is related to differences in the fields nomenclature. For instance, relationship status includes many more options in the Facebook Social Network data than in Census data, such as `engaged' and `in a domestic partnership'. For education attainment, in particular, we need to group different categories from Census data, since they provide separate categories for people between 18 and 24 years old and above 25\footnote{\url{https://data.census.gov/cedsci/table?g=0100000US&tid=ACSST5Y2017.S1501&q=S1501%https://factfinder.census.gov/bkmk/table/1.0/en/ACS/17_5YR/S1501/0100000US
}}. Finally, age is limited on Facebook since the platform only allows users above 13 years old. Table \ref{table:education_level_mapping} details the education attainment fields of the Census and Facebook used to compose the total audience in each category.
\begin{table*}[tb]
\centering
\begin{tabular}{ |p{2.5cm}|p{8cm}|p{6cm}|}

\hline
	 \textbf{Category} & \textbf{Census} & \textbf{Facebook} \\ \hline
	Incomplete High School & Less than high school graduate (18-24), Less than 9th grade (above 25), 9th to 12th grade, no diploma (above 25) & In high school,Some high school \\ \hline
	High School & High school graduate -includes equivalency (18-24), High school graduate - includes equivalency (above 25) & High school grad \\ \hline
	Some College & Some college, no degree (above 25) & In college, Some college \\ \hline
	College & Associate's degree (above 25), Bachelor's degree (above 25), Some college or associate's degree (18-24), Bachelor's degree or higher(18-24) & College grad\\ \hline
	Grad Degree & Graduate or professional degree (above 25) & Some grad school, Master degree, Doctorate degree, Professional degree, Studying grad school \\ \hline

\end{tabular}
\caption{\textbf{Educational attainment mapping.}}
\label{table:education_level_mapping}
\end{table*} 

Another issue is related to small-sized targeted populations in Facebook. For subpopulations smaller than one thousand users, the Facebook advertising platform returns the value $1000$. This is a mechanism to prevent advertisers to succeed in unveiling the identity of a certain user by creating a target formula that leads to a unique user. As we focused on the most populous cities, this limitation represented no problem in our study. However, this privacy protection mechanism could represent a limitation for obtaining demographic data from Facebook in small cities. Finally, we are not able to account for fake information about gender, age, or level of education provided by users in their public profile.

\section{Analysis}

In this section, we aim at comparing the demographic distributions collected through the Facebook Ads with consolidated offline results. For most of the validation in this current study, we used recent baselines provided by the Census Bureau estimation studies. 

Firstly, we characterize the distribution of selected demographic attributes in the U.S. as a whole. In a second analysis, we dig into states and cities to check the demographic distribution of the Facebook population with a fine-grained perspective. 
Finally, we present a report about immigrants in the U.S..


\subsection{Country-level analysis}

\begin{figure}[tb]
    \centering
    \includegraphics [width=8.5cm,height=\textheight, keepaspectratio ]{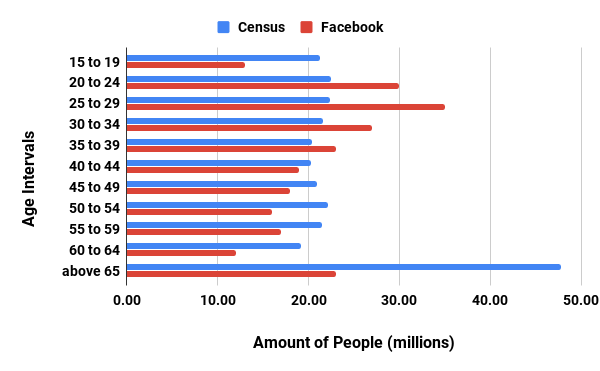}
    \caption{Population grouped by age}
    \label{fig:population_by_age}
\end{figure}

Facebook registers 230 million active users who live in the U.S. (July 2018). Figure \ref{fig:population_by_age} shows the population size by age groups. Not surprisingly, the Facebook population sizes for people under 19 and above 65 are significantly lower than the real U.S. population provided by Census data. This may be explained because the younger group does not include people under 13 since Facebook does not allow children to register. In spite of increasing their participation in Social Networks in the last years, people above 65 years old are, in general, less inclined to use OSNs than young people as also highlighted by related research~\cite{gil2019demographic}.

In opposition to these underrepresented groups, Facebook overestimates the population with ages between 20 to 39 years old in comparison with the Census. This large population of adults raised some criticism about the way Facebook calculates its audience size, and some suggested that Facebook might be inflating the numbers in order to increase their revenue \footnote{\url{www.dataiq.co.uk/article/news-analysis-facebook-v-census-out-count}}
\footnote{\url{www.businessinsider.com/facebook-tells-advertisers-reaches-25-million-more-people-than-exist-us-census-data-2017-9}}. Facebook alleged in a statement that ``Reach estimations are based on a number of factors, including Facebook user behaviors, user demographics, location data from devices, and other factors. They are designed to estimate how many people in a given area are eligible to see an ad a business might run. They are not designed to match population or census estimates. We are always working to improve our estimates''. It is possible that these inflated numbers may be the result of people having multiple accounts, including potentially business accounts.

\begin{figure*}[t!]
\centering
    \begin{subfigure}[b]{0.9\columnwidth}
    \centering
  \includegraphics[width=\columnwidth]{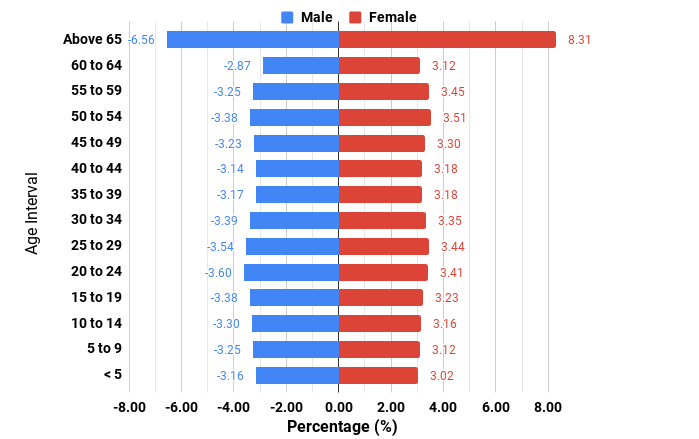}
    \caption{Census}
    \end{subfigure}\hfill
  \begin{subfigure}[b]{0.9\columnwidth}
    \centering
  \includegraphics[width=\columnwidth]{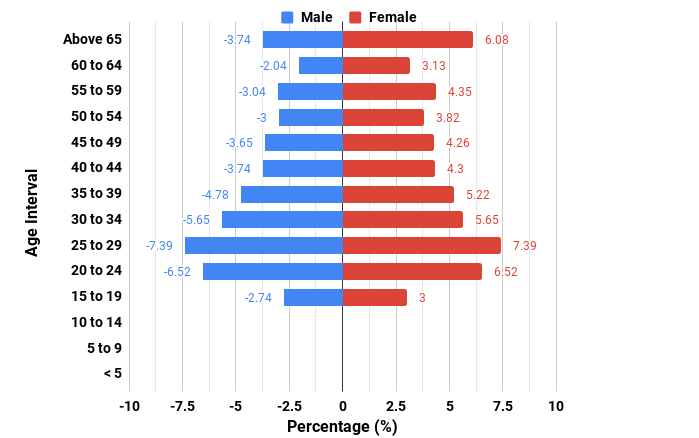}
    \caption{Facebook}
    \end{subfigure}\hfill
     \caption{\textbf{Age distribution by gender.}}
    \label{fig:us_population_age_pyramid}
\end{figure*}

Figure \ref{fig:us_population_age_pyramid} depicts the age distribution  by gender in a pyramid bar chart. We grouped all the intervals with persons older than 65 in the above 65 bar since Facebook does not allow stratifying users above 65 years old more accurately. Additionally, the Facebook bar chart does not contain information for the population under 15, since Facebook does not allow users younger than 13. As shown in the previous age distribution figure, the more represented interval on Facebook ranges from 20 to 39 years old. Curiously, the number of male and female users are exactly the same for the most populous three ranges: 20 to 24 years old (6.52\%), 25 to 29 (7.39\%) and 30 to 34 years old (5.65\%). The figure also shows that women are overrepresented in the older intervals, especially, above 65 years old, in which the number of women is 62\% bigger than men. This difference is only 24\% in the Census distribution. 

The overall gender distribution of Facebook is slightly biased towards women. While men comprise 49.2\% of the United States population and women account for 50.8\% in the ACS survey, the women population on Facebook is 52.8\%.

\begin{figure*}[t!]
\centering
    \begin{subfigure}[b]{0.9\columnwidth}
    \centering
  \includegraphics[width=\columnwidth]{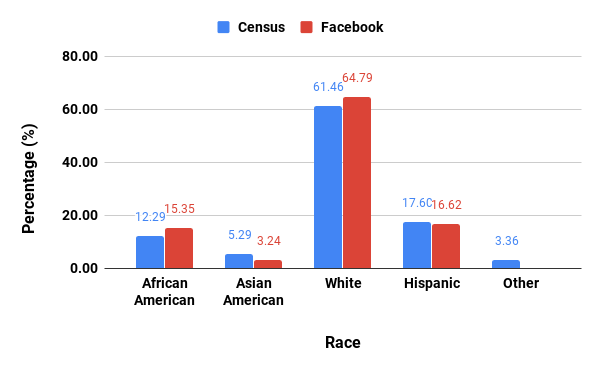}
    \caption{Percent values}
    \end{subfigure}\hfill
  \begin{subfigure}[b]{0.9\columnwidth}
    \centering
  \includegraphics[width=\columnwidth]{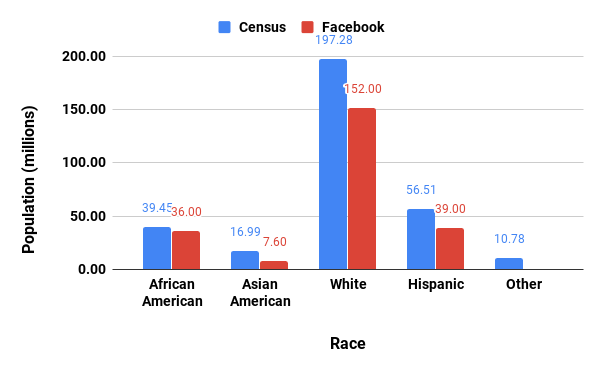}
    \caption{Raw values}
    \end{subfigure}\hfill
     \caption{\textbf{Racial and ethnic distributions in the U.S., together with proxies for these quantities in Facebook.}}
    \label{fig:population_by_race}
\end{figure*}

Figure \ref{fig:population_by_race} compares the distribution of the U.S. population in terms of races and ethnicity. The Facebook Marketing API includes an attribute called affinities that identifies affinity of  users to groups like Hispanics, African-Americans and Asian-Americans. These can only be considered as rough approximations to the definitions used in the actual Census. We got estimates of the audience size for each of these categories and then considered the remainder as non-Hispanic whites (we refer to them as whites for simplicity). We noticed that the distribution of races across Facebook is quite similar to the Census distributions, being slightly over-represented by African-Americans and whites, and underrepresented by Asian-American and non-white Hispanics (see Figure \ref{fig:population_by_race} (a)). When analyzing raw values, depicted in Figure \ref{fig:population_by_race} (b) we can check that the over-representation found in the age intervals category is not observed in the race distribution, at least not directly. The African-American population on Facebook is only 3 million less than the African-American population in the Census. Considering that the Facebook population includes only users above 13 years old, the 36 million population of African-Americans may lack at least 10 million of African-American under 13 years old, which would also characterize an over-representation of this particular ethnicity.

\begin{figure*}[t!]
\centering
    \begin{subfigure}[b]{0.9\columnwidth}
    \centering
  \includegraphics[width=\columnwidth]{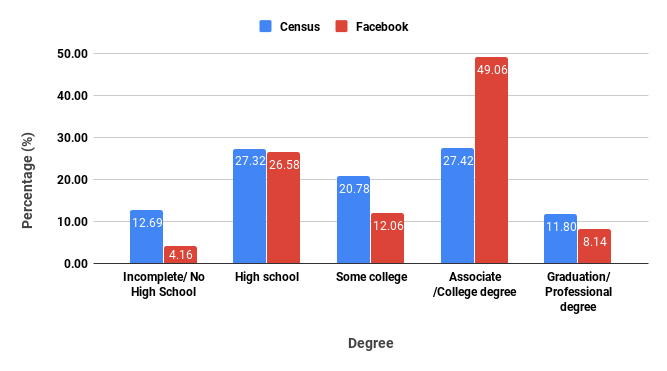}
    \caption{Percent values}
    \end{subfigure}\hfill
  \begin{subfigure}[b]{0.9\columnwidth}
    \centering
  \includegraphics[width=\columnwidth]{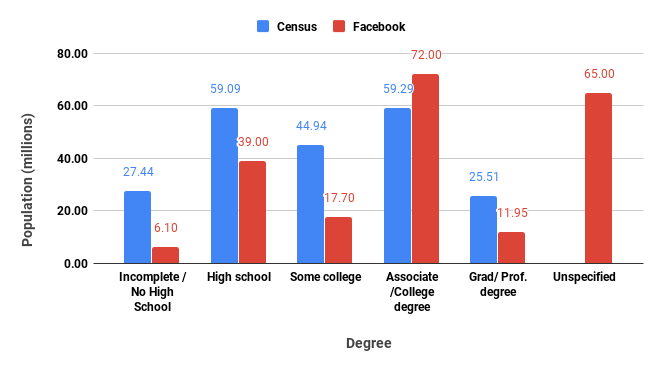}
    \caption{Raw values}
    \end{subfigure}\hfill
     \caption{\textbf{Education level distribution in the U.S..}}
    \label{fig:us_population_by_education_level}
\end{figure*}

An important challenge when considering data produced by OSN users is that there is no guarantee the information is correct. In many cases, users insert information in their profiles to mock some situation or subject and sometimes they include some information to avoid leaving the field blank. Creation of fictitious job titles or colleges may be found with relative frequency. Another situation occurs when the users do not fill out their public profile due to privacy concerns or simply do not wish to spend their time doing this. The education level field, for instance, is not filled out by 65 million users as can be seen in figure \ref{fig:us_population_by_education_level} (b). This figure depicts the educational attainment in the U.S.. Note that the number of people with the associate or college degree on Facebook also overpasses the amount informed by the census authority. The percentages are depicted in figure \ref{fig:us_population_by_education_level} (a).

In terms of income level, data obtained from Facebook partially differs from the Census. Firstly, Facebook only infers the income with values above 30 thousand dollars a year. Another observation is that the Facebook population is much richer than the real population with an overestimation of the number of people who earn more than 50 thousand dollars. The income level is provided by one of the Facebook partners that help the OSN to provide more detailed targeting options to advertisers, especially regarding the purchasing and offline behavior. However, data provided by some of these partners is no longer available since October 2018\footnote{\url{https://about.fb.com/news/h/shutting-down-partner-categories/
%https://newsroom.fb.com/news/h/shutting-down-partner-categories/
}}. It is not clear how Facebook and partners classify the users by the amount of money they earn, but the bias toward the richer, again, may raise some criticism on Facebook, since it would inflate the audiences most targeted by advertisers. As the baseline for income level, we considered full-time, year-round workers with earnings in the Census table named ACS\_17\_5YR\_S2001\footnote{\url{https://data.census.gov/cedsci/table?g=0100000US&tid=ACSST5Y2017.S2001&q=S2001}
}.


\begin{figure*}[t!]
\centering
    \begin{subfigure}[b]{0.9\columnwidth}
    \centering
  \includegraphics[width=\columnwidth]{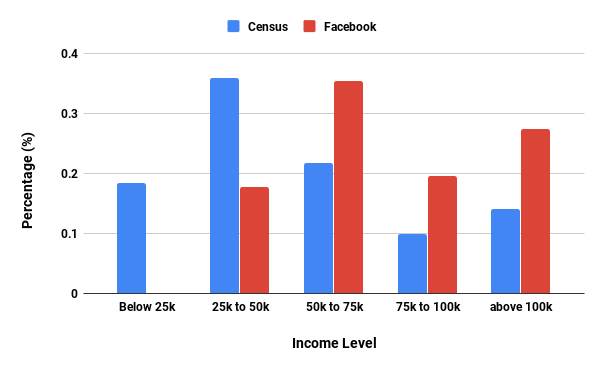}
    \caption{Percent values}
    \end{subfigure}\hfill
  \begin{subfigure}[b]{0.9\columnwidth}
    \centering
  \includegraphics[width=\columnwidth]{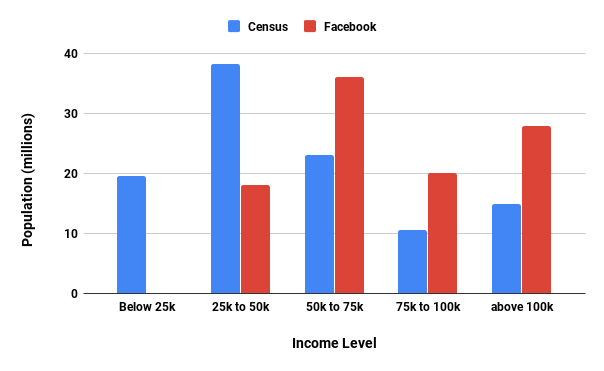}
    \caption{Raw values}
    \end{subfigure}\hfill
     \caption{\textbf{Income level distribution in the U.S..}}
    \label{fig:us_income_level}
\end{figure*}


\subsection{Finer granularity - states and cities}

\begin{figure}[tb]
    \centering
    \includegraphics [width=8.5cm,height=\textheight, keepaspectratio ]{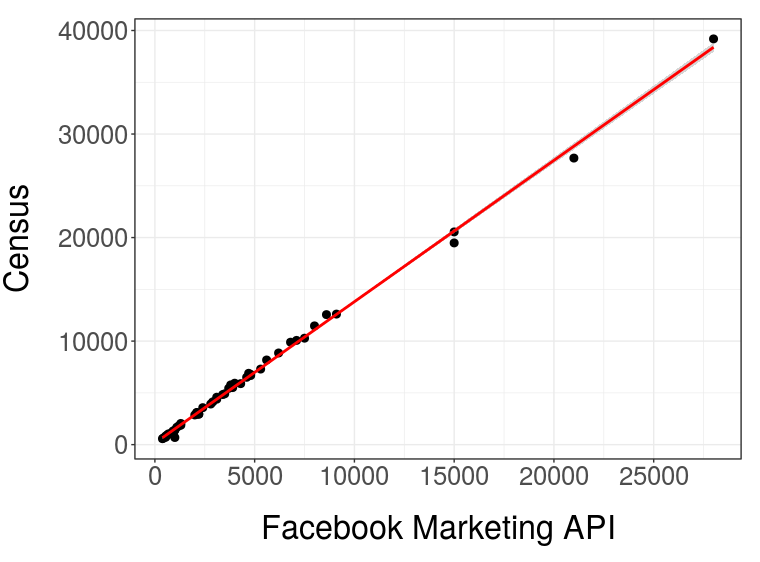}
    \caption{Population by state}
    \label{fig:population_by_state}
\end{figure}

The next analysis aims at checking if the demographic data obtained with the Facebook Marketing API captures the variation across different locations. Firstly, we compared the total population in each one of the 50 U.S. states and  D.C. according to Facebook Ads and Census, and we find a very high Pearson correlation ($0.9988$) (see figure \ref{fig:population_by_state}). The District of Columbia has the highest proportion of the population on Facebook compared to the Census population with a higher population on Facebook than in the real world, one million on Facebook compared to the less than $700$ thousand official estimate. This may be due to border characteristics of the U.S. capital that lead to a misleading inference of location from Facebook. Apart from the U.S. capital, the states with the highest proportion of population on Facebook are New York with 76\%, and Alaska and Texas with 75\%, whereas the states less represented online are New Mexico with 64\% and South Dakota with 65\%. 

In a second analysis, we compared the fluctuation of race, income level, political leaning, and educational attainment across all the 50 U.S. states and capital again. Table \ref{table:corr_categories_usa_states} (state level) synthesizes how correlated the data collected with our framework are when compared with data from the ACS across states, by calculating Pearson correlation with 95\% confidence intervals (CI). Notice that, in terms of race, the correlation is very high for African-Americans, Asian-Americans, and Hispanics, which means that Facebook accurately infers the origins of a user. Recall that Facebook does not classify white people, and for the calculation of this attribute, we excluded the other three races, which may explain, in part, the lower accuracy in this case. 

\begin{figure*}[t!]
\centering
    \begin{subfigure}[b]{0.6\columnwidth}
    \centering
  \includegraphics[width=\columnwidth]{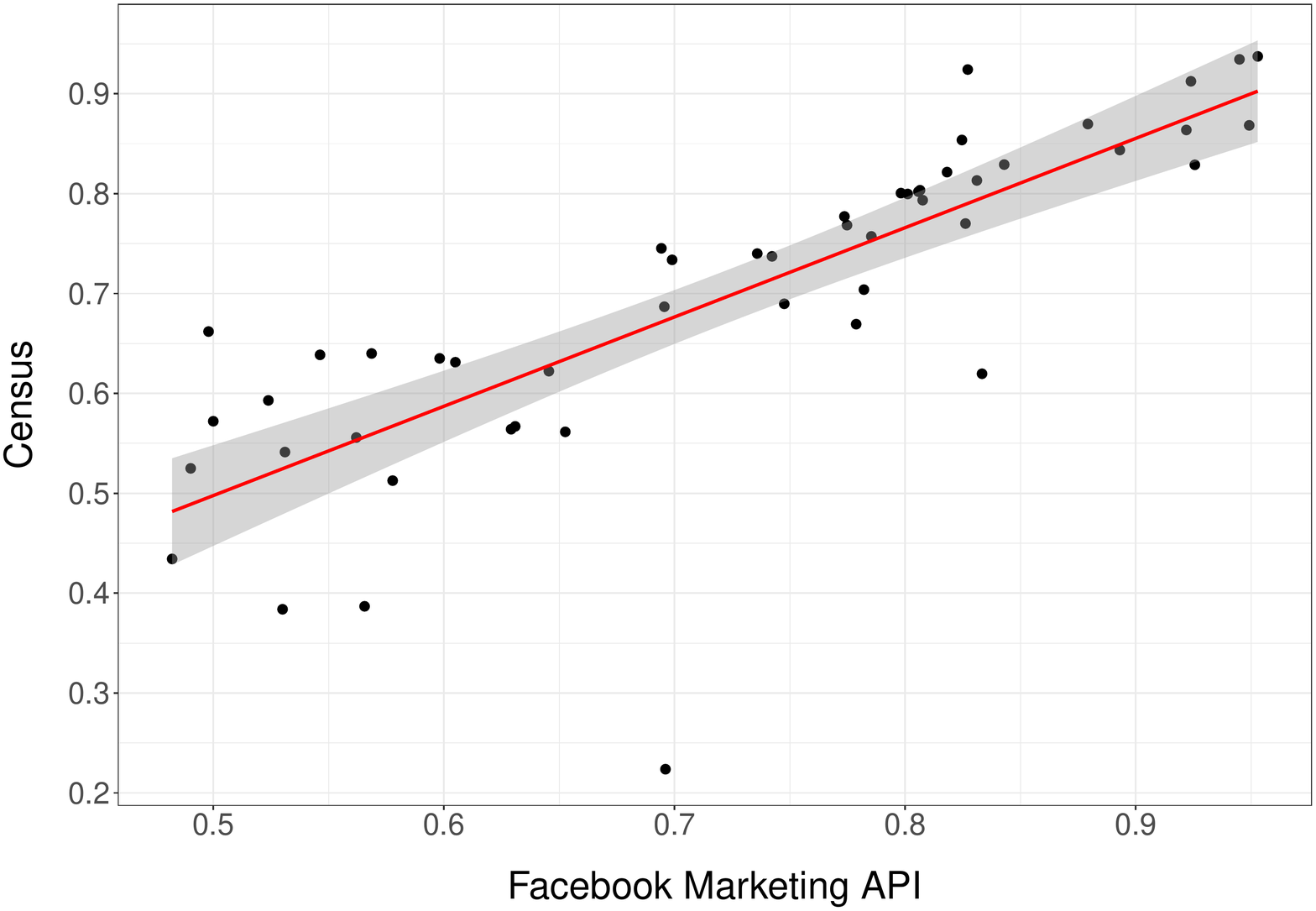}
    \caption{Race - White}
    \end{subfigure}\hfill
  \begin{subfigure}[b]{0.6\columnwidth}
    \centering
  \includegraphics[width=\columnwidth]{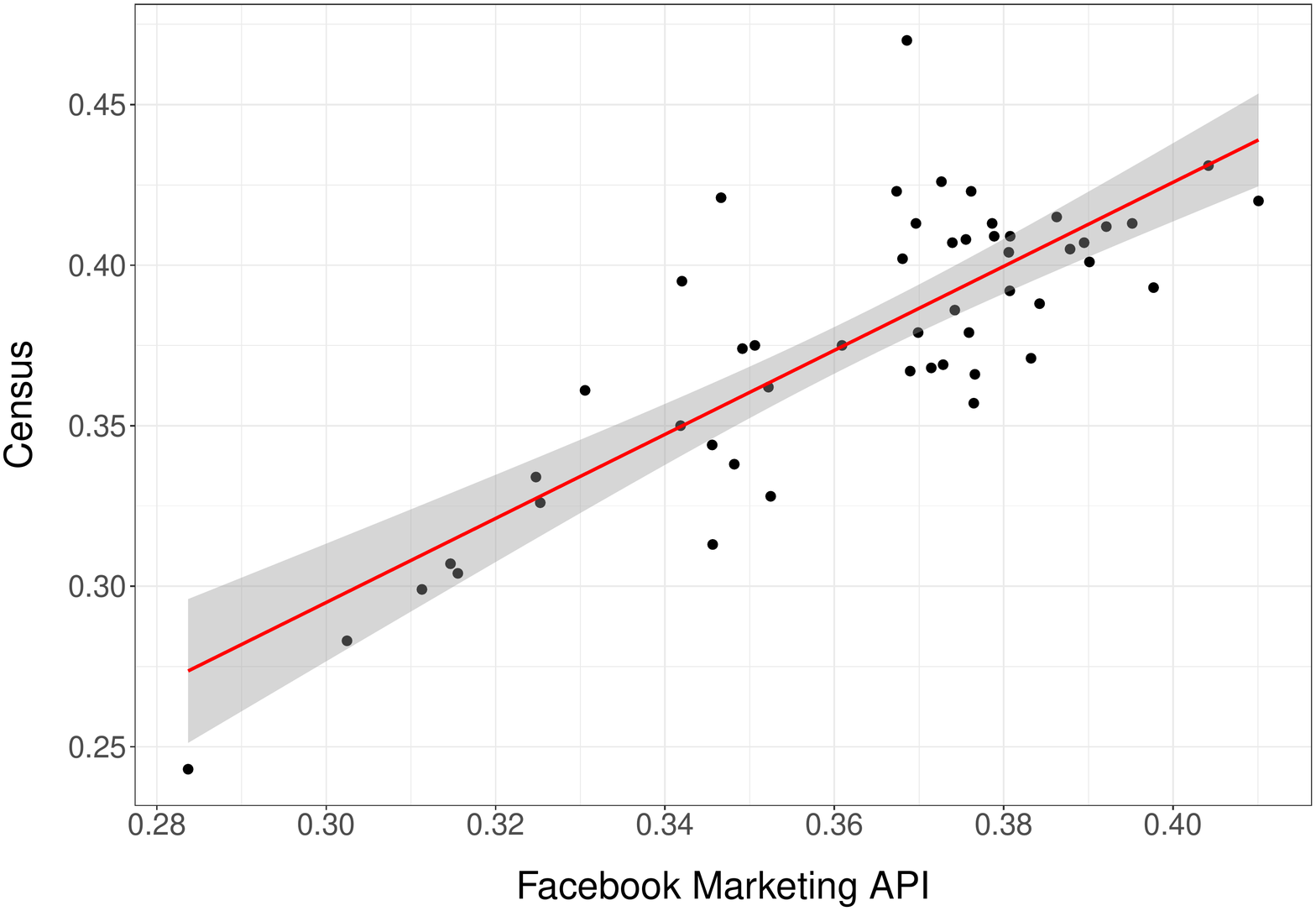}
    \caption{Income - 50k to 75k(*)}
    \end{subfigure}\hfill
  \begin{subfigure}[b]{0.6\columnwidth}
    \centering
  \includegraphics[width=\columnwidth]{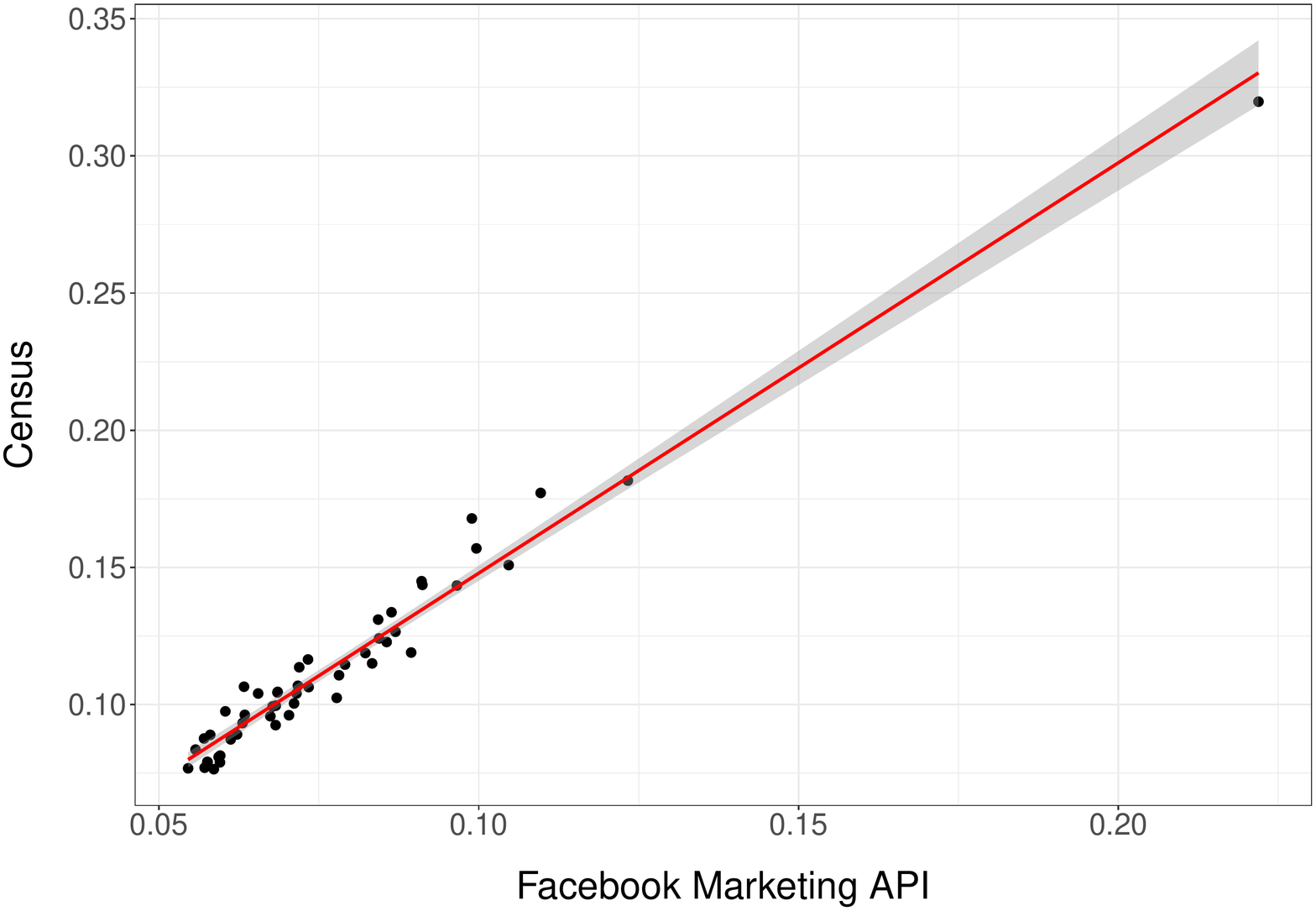}
    \caption{Education Level - Grad School}
    \end{subfigure}\hfill    
     \caption{\textbf{Comparing the percentage of selected attributes in data from Facebook Ads and from the Census Bureau, for U.S. states.}}
    \label{fig:us_attributes_across_states}
\end{figure*}

Figure \ref{fig:us_attributes_across_states} (a) plots the correlation for the white population across states. By analyzing the largest discrepancies, we observed that they include Hawaii and Alaska, both states with particular ethnic groups calculated by  the Census, but not assessed on Facebook: Native Hawaiians, and Alaska Natives. For Hawaii, by using our framework we found almost 70\% of white people whereas there are less than 23\% according to Census data. The difference for Alaska  is about 21\% (83\% with our framework rather than 62\% with Census). Alabama is the third state with the highest difference, less than 50\% on Facebook against 66\% in Census data. 

In terms of income level, the best correlation is for people with high earnings (above 100k dollars per year). For other levels, we find a poor correlation, except for an interesting observation regarding this particular attribute: the percentage of Facebook users with annual income level between 50k and 75k dollars are highly correlated with the group 25k to 50k across states in the Census  data (see 50k to 75k (*) in table \ref{table:corr_categories_usa_states} and figure \ref{fig:us_attributes_across_states} (b)).

By checking the educational attainment rows in table \ref{table:corr_categories_usa_states} we find a similar result, only two attributes presented a high correlation, including people with high school degree and with a graduation degree, especially the most educated people. Figure \ref{fig:us_attributes_across_states} depicts the high correlation for the graduate school education level. This suggests that when a Facebook user fills out its education level with some graduate school it more likely to be correct, compared to college graduates. 

The last attribute checked across all states was political leaning, for which we found a high correlation for left leaning and right leaning and a poor correlation for moderates. The lower correlation for moderates  may be explained by the baseline we used, that is based on annual state averages of party affiliation from Gallup Daily tracking. This data set is not ideal to detect the proportion of moderates in each state. 

\begin{table*}[tb]
\centering
\begin{tabular}{ | c | l | l | l | l | l | }
\hline
	\multirow{2}{*}{\textbf{Category}} & \multirow{2}{*}{\textbf{Dimension}} & \multicolumn{2}{|c|}{\textbf{State Level}} & \multicolumn{2}{|c|}{\textbf{City Level}}  \\ \cline{3-6}
	& & \textbf{Pearson C.} & \textbf{CI (95\%)} & \textbf{Pearson C.} & \textbf{CI (95\%)}  \\ \hline
	\multirow{4}{*}{\textbf{Race}}& African-American & 0.97 & [0.95,0.98] & 0.94 & [0.90,0.97]\\ \cline{2-6}
	 & Asian-American & 0.97 & [0.95,0.98] & 0.94 & [0.89,0.96] \\ \cline{2-6}
	& Hispanic & 0.97 & [0.95,0.98] & 0.96 & [0.94,0.98]   \\ \cline{2-6}
	& White & 0.82 & [0.71,0.90] & 0.86 & [0.77,0.92]\\ \hline
	\multirow{5}{*}{\textbf{Income Level}}& 25k to 50k & 0.76 & [0.62,0.86] & 0.69 & [0.51,0.81]  \\ \cline{2-6}
	 & 50k to 75k & -0.33 & [-0.55,0.06] & -0.13 & [-0.39,0.16]    \\ \cline{2-6}
	 & 50k to 75k (*) & 0.83 & [0.72,0.90] & 0.73 & [0.56,0.84] \\ \cline{2-6}
& 75k to 100k & 0.67 & [0.49,0.80] & 0.55 & [0.31,0.72]   \\ \cline{2-6  }
	 & above 100k & 0.93 & [0.88,0.96] & 0.83 & [0.72,0.90] \\ \hline
	\multirow{3}{*}{\textbf{Educational Attainment}} & Incomplete High School & 0.34 & [0.08,0.57] &  0.36 & [0.09,0.58] \\ \cline{2-6}
	 & High School & 0.87 & [0.77,0.92] &  0.71 & [0.54,0.83] \\ \cline{2-6}
	 & Some College & 0.55 & [0.32,0.71] &  0.51 & [0.27,0.69] \\ \cline{2-6}
	 & College & 0.62 & [0.41,0.76] &  0.57 & [0.35,0.73]     \\ \cline{2-6}
	 & Grad School & 0.98 & [0.97,0.99] &  0.86& [0.77,0.92]  \\ \hline
	\multirow{3}{*}{\textbf{Political Leaning}} & Left leaning & 0.87 & [0.79,0.93] & - & - \\ \cline{2-6}
	 & Moderate & 0.02 & [-0.26,0.29] & - & - \\ \cline{2-6}
	 & Right leaning & 0.91 & [0.85,0.95] & - & - \\ \hline	 
\end{tabular}
\caption{\textbf{Correlations for demographic categories across U.S. states and cities.}}
\label{table:corr_categories_usa_states}
\end{table*}

In order to provide a comparison at a more fine-grained level, we conducted similar comparisons for the 50 most populous cities in the U.S. with results presented in table \ref{table:corr_categories_usa_states} (city level). 
It also presents the Pearson correlation as well as the values with 95\% confidence intervals for each category, except for political leaning. The last one was not included due to the lack of available information at that level of granularity. We observed that the correlation is often a little lower than at the state level. 

One of the factors that may explain the lower correlation in compared to the state-level analysis is related to the Facebook data collection process. When selecting the city on the Facebook advertising platform, we must define the name of the city and the radius of the collection that limits the population included in the target audience. The default radius is 30 miles and the lowest radius available is 10 miles. In our collection, we used the 10-mile radius option, which does not match the official borders of the city, meaning that the calculated demographics may include users from neighboring regions or exclude users that were supposed to be included in the audience. The Census population of Arlington, in Texas, next to Fort Worth (a large city with 874 thousand inhabitants) is roughly 390 thousand people (ACS 2017) whereas the population on Facebook is 1 million. The same issue holds for Minneapolis (neighbor to the large city of St. Paul, state capital) with a 411 thousand population according to the Census and 1.1 million as counted by Facebook. In both cases, the final audience includes people from outside the city borders. Conversely, for New York City, the population size is similar in both measurements, 8.5 million people.    

\subsection{U.S. immigrants analysis}

In this analysis, we compared the population size of immigrants in the U.S.. We used the table B05006 from the ACS 5 Year Estimates as the baseline. For the Facebook data, we collected the number of immigrants for all available countries on the platform. 

\begin{figure*}[t!]
\centering
    \begin{subfigure}[b]{0.9\columnwidth}
    \centering
  \includegraphics[width=\columnwidth]{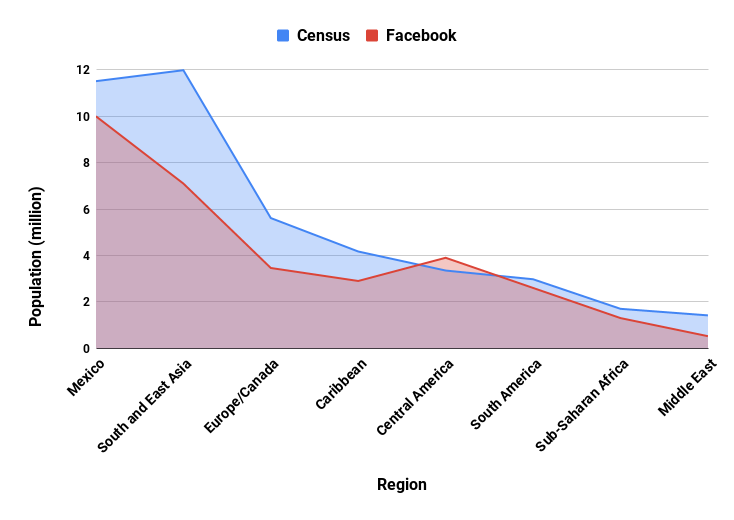}
    \caption{Regions}
    \end{subfigure}\hfill
  \begin{subfigure}[b]{0.99\columnwidth}
    \centering
  \includegraphics[width=\columnwidth]{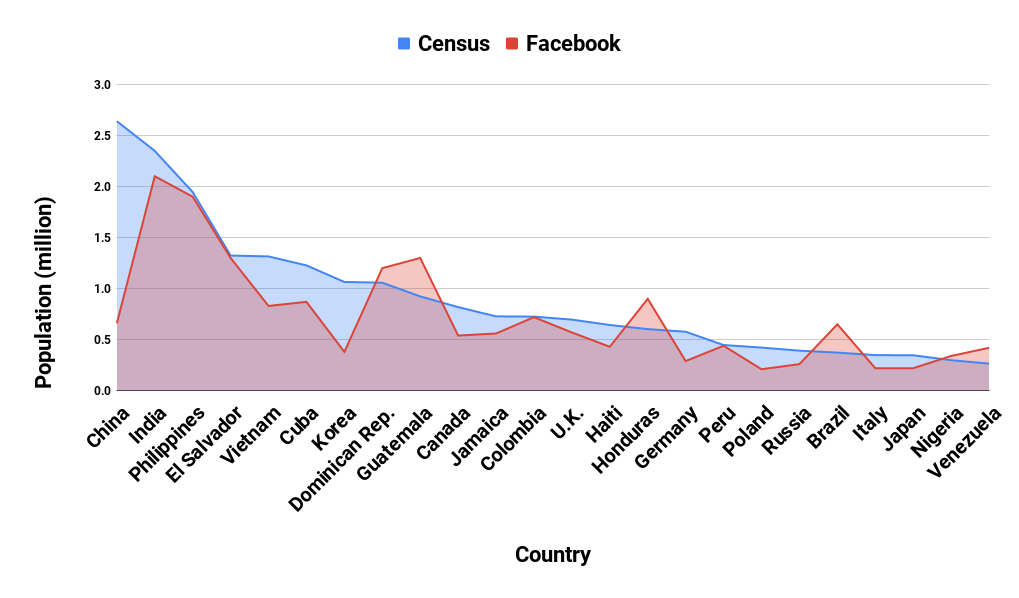}
    \caption{Top 25 countries with more immigrants (except Mexico)}
    \end{subfigure}\hfill
     \caption{\textbf{Population of immigrants to the U.S. by region and country of origin.}}
    \label{fig:expats_origin}
\end{figure*}

Figure \ref{fig:expats_origin} (a) depicts the number of immigrants living in the U.S. with origins in different regions around the world. Notice that the population size on Facebook is smaller than in the Census data for all regions of origin except for Central America, for which  the Facebook population is nearly 550 thousand larger than the Census. There are different gaps between both measurements for the other regions. For immigrants from South and East Asia, for instance, the Census population size is roughly 4.8 million larger than the Facebook population. This may be explained by the banishment of Facebook from China, meaning that the largest OSN is not the best platform to remain in contact with compatriots that still live in the origin country. On the other hand, the gap in South America is small, with 2.6 million immigrants according to Facebook and 2.9 millions according to Census data.  

Figure \ref{fig:expats_origin} (b) allows us to check the difference in the country level for the top 25 countries with more immigrants in the U.S.. We can notice, for instance, that the gap is huge for Chinese Facebook users, 2.64 millions according to the Census and only 0.66 millions on Facebook. On the other hand, Central and South American countries such as Guatemala, Honduras, Brazil, and Venezuela have more immigrants on Facebook than those calculated by the Census. In other examples for the same origin region, the numbers are very similar in both measurements, for example, El Salvador, Dominican Republic, and Peru. This finding might indicate that the Census is underestimating the population of immigrants with origins in specific countries. For the sake of simplicity, Mexico was included in the regions' figure and not in the top 25 countries due to the high number of immigrants from this country. 

We should mention that we were not able to count the immigrants from some particular countries on Facebook and they were excluded from our analysis. Those countries excluded from the top  25 list are Iran, Pakistan, Ukraine, and Ecuador. For the regions figure, we were not able to include a considerable number of countries due to the absence of information about these countries on Facebook. The percentage of missing countries per region are the following: South and East Asia (48\%), Europe (31\%), Caribbean (83\%), Central America (43\%), South America (50\%), Middle East (60\%) and Sub-Saharan Africa (69\%).

\subsection{Correction Factors}

In addition to the Census and Facebook distributions for each demographic attribute, we also computed a correction factor that allows one to multiply it by the Facebook distribution to obtain the Census distribution as a result.

The correction factors, computed for each demographic dimension and for all levels (country, state and city level) can be very useful for demographic research. One particular use is deriving the actual population for some distribution of interest calculated previously through the Facebook advertising platform. Suppose someone wants to identify how many people are interested in an activity, brand or any other entity in a particular geographic region, stratified by gender. One can collect the distribution in the Facebook advertising platform (by manually selecting the audiences on the ad creator graphic interface) and derive the population interested in that entity after multiplying the numbers by the appropriate correction factor, that is intended to adjust the estimates for known biases. Recall that the audience estimation does not require the publication of an ad and does not incur any expense. Facebook provides more than 250 thousand attributes~\citep{speicher-2018-targeted} that can be used to select a huge range of audiences that can be further extrapolated to the real world.   

In addition to the statistical value, there is also a sociological value associated to the corrections factors. They enable researchers to assess which  groups are over- and under-represented in the online world. It is widely known that certain groups are more or less represented on Facebook, and by using the correction factors, we can quantify this bias. Table \ref{table:correction_factor} shows the percentage of African-American measured by Facebook and the Census as well as the correction factor (CF), for six U.S. states. Notice that, the Census value can be obtained by multiplying the CF by the Facebook value, indicating that the lower the CF the less under-represented (or more over-represented) the Facebook users are. The top three rows show the three most over-represented states on Facebook with respect to  this demographic dimension, whereas the bottom rows present the states under-represented for this variable. For this demographic dimension, African-Americans, Facebook is over-represented in 48 out of 51 states.

\begin{table*}[tb]
\centering
\begin{tabular}{ | l | l | l | l | }
\hline
	\textbf{US State} & \textbf{\% Facebook} & \textbf{\% Census} & \textbf{CF} \\ \hline
	\textbf{West Virginia} & 14.061 & 3.507 &  0.24939   \\\hline
	 \textbf{Montana}& 1.256 & 0.396 & 0.31546   \\ \hline
	 \textbf{Hawaii}& 4.216 & 1.687 & 0.40007    \\ \hline
	\textbf{District of Columbia} & 46.829 & 46.871 &  1.0009  \\ \hline
	 \textbf{Massachusetts}& 6.598 & 6.682 & 1.01279  \\ \hline
	\textbf{South Dakota} & 1.495 & 1.671 & 1.11739   \\ \hline	
\end{tabular}
\caption{\textbf{Correction factors for the `African-American' dimension (most biased states are shown).}}
\label{table:correction_factor}
\end{table*}

\section{Concluding Discussion}

In this work, we leveraged the Facebook advertising platforms to compile the size of demographic groups of Facebook users in the U.S., along seven different attributes: gender, race, age, income, education level, political leaning, and country of origin for immigrants. We calculate the demographic distributions at  different levels of granularity: country, state, and city level.

We analyzed the Facebook Census by comparing it with official data provided by the Census Bureau and estimates offered by Gallup. We confirmed the observation of a bias in the online population towards young people and women. We also verified that the distributions of race and ethnicity, in particular, are fairly similar to the real distributions at all levels of granularity. The education level obtained online seems to be over-represented for the college degree level. However, for high school and grad school degree, we observe similar distributions compared to the offline data at the state level. The same occurs for income level:  Facebook values for the higher income levels (above 100k per year) are fairly close to what is provided by  Census data. We also assessed that the Facebook values for educational attainment and income level at the city level are not as good as data at the state level. This is in part related to the issues of identifying city borders. In terms of immigration, the online data seem to follow the same tendency of official data, except for immigrants from South America and Central America for whom Facebook data tend to be higher than Census data.This may indicate issues of under-estimation in official sources. Finally, with respect to political leaning, Facebook  provides accurate distributions at state level for conservative and liberal people, but not so much for moderates. 

Our methodology showed to be valuable as it clarifies the dimensions for which Facebook data are closer to the actual population estimates, as well as details about  biases across several dimensions. As a matter of fact, we calculated correction factors for each dimension at each level of granularity for which we had data. Our correction factors  could be recomputed periodically as biases may change over time. This information could be used to generate bias-adjusted population estimates for various dimensions and needs, in a timely manner.

As a final contribution, we release our estimates and correction factors. We expect that our data set and approaches can open many avenues of research, especially for those interested in understanding how biases in the population of Facebook users are changing over time.

\begin{acks}
This research was partially supported by Conselho Nacional de Desenvolvimento Cient{\'i}fico e Tecnol{\'o}gico (CNPq), and Coordena{\c c}{\~a}o de Aperfei{\c c}oamento de Pessoal de N{\'i}vel Superior (CAPES).
\end{acks}
\bibliographystyle{ACM-Reference-Format}
\bibliography{references.bib}

\end{document}